# Anapole, chiral and orbital states in $Mn_3Si_2Te_6$

S. W. Lovesey[1, 2, 3]

[1] ISIS Facility, STFC, Didcot, Oxfordshire OX11 0QX, UK

[2] Diamond Light Source, Harwell Science and Innovation Campus, Didcot, Oxfordshire OX11 0DE, UK

[3] Department of Physics, Oxford University, Oxford OX1 3PU, UK

**Abstract** The ferrimagnet $Mn_3Si_2Te_6$ attracts attention because of a recently discovered colossal magnetoresistance (CMR) with unique magnetic field properties. An improved magnetic structure for the material has emerged from a neutron diffraction study linked to understanding the CMR. A deeper theoretical investigation of the magnetic structure has now revealed anapole, chiral and orbital states of manganese ions not previously mentioned. Moreover, it is shown that existence of these states in the low temperature form of $Mn_3Si_2Te_6$, with a magnetic field applied, can be tested by neutron and resonant x-ray diffraction.

## I. INTRODUCTION

Silicon manganese telluride ($Mn_3Si_2Te_6$) is semiconducting and ferrimagnetic below a temperature $\approx$ 78 K [1]. A trigonal structure for the compound was established many years before recent measurements of its colossal magnetoresistance (CMR) and an improved ferrimagnetic structure [2, 3, 4, 5]. The nature of CMR in $Mn_3Si_2Te_6$ sets it aside from other CMR materials. Basal plane conductivity drastically increases (a factor $10^7$) when an external magnetic field is applied along the magnetic hard axis normal to the plane, and there is only a modest change to the conductivity when the field is applied parallel to an easy axis in the plane [6, 7, 8].

The theoretical study reported here paves the way for numerous tests of the improved magnetic structure of $Mn_3Si_2Te_6$. A tried and tested structure is required for meaningful interpretations of experimental data and a consensus understanding of the unusual magnetic properties [5]. Tests can result in a refinement or rejection of the defining magnetic space group, which is a standard procedure in establishing a reliable space group for a chemical structure. The study takes the form of calculations informed by magnetic symmetry of scattering amplitudes for beams of neutrons and x-rays. Bragg diffraction of these radiations unveils anapole, chiral and orbital states of a crystal. Calculations predict all these states for $Mn_3Si_2Te_6$ in an applied magnetic field, none of which have been mentioned in the literature, to the best of our knowledge. Anapoles (a Dirac dipole depicted in Fig. 1) appear in diffraction amplitudes of neutrons, and x-rays tuned to an atomic resonance [9 - 12]. Unlike loopy entities [8, 13], axial and Dirac atomic multipoles have specific discrete symmetries. Tuning the energy of x-rays to an atomic resonance has two obvious benefits in diffraction experiments. In the first place, there is a welcome enhancement of Bragg spot intensities and, secondly, spots are element specific [14, 15]. Another attribute of resonant x-ray Bragg diffraction is realized with circularly polarized x-rays that detect charge-like and magnetic chiral states [16, 17, 18]. Absorption of x rays at the K edge exposes orbital states alone [19], while spin and orbital degrees of freedom contribute to Bragg spots enhanced by absorption at L edges, for example

[20]. Some iridates and ceramic superconductors harbour Dirac multipoles that have been shown to diffract neutrons [21]. Notably, the existence of Dirac quadrupoles in the pseudo-gap phases of cuprate superconductors YBCO and Hg1201 account for magnetic Bragg diffraction patterns [22, 23]. In the present theoretical study of neutron and x-ray diffraction by $Mn_3Si_2Te_6$ particular attention is given to weak magnetic Bragg spots that do not appear in the diffraction pattern of the parent structure. The intensity of one such basis-forbidden Bragg spot for $Mn_3Si_2Te_6$ at a temperature ≈ 5 K as a function of an applied magnetic field has been reported [5]. By their nature, basis-forbidden Bragg spots reveal details of an electronic structure. Best known, perhaps, is Templeton-Templeton (T & T) scattering due to departures from spherical symmetry in atomic charge distributions [24, 25]. According to Neumann's Principle, such departures delineate the local symmetry of resonant ions [26, 27, 28]. As we mentioned, analogous effects occur in magnetic diffraction that is the principal subject of our study. The two independent Mn ions in $Mn_3Si_2Te_6$ have different magnetic properties on account of their different environments.

The nominal atomic configuration of a manganese ion in $Mn_3Si_2Te_6$ is $3d^5$ ($Mn^{2+}$). A pure configuration $3d^5$ allows axial magnetic dipoles alone, with spin = 5/2 and no orbital angular momentum. Silicon manganese telluride shows a surprisingly large magnetic anisotropy. In simulations of its magnetic structure, the anisotropy field originates from an anisotropy in the Mn orbital moments, and concomitant orbital angular momentum [3]. An admixture of states with opposite parities creates Dirac multipoles, e.g., d and p atomic orbitals, and informative model calculations for V and Cu ions in $V_2O_3$ and CuO, respectively, appear in Refs. [11, 29, 30]. By the same token, simulations of parity-odd electric dipole- electric quadrupole (E1-E2) x-ray scattering amplitudes depend on an admixture of states with opposite parities [31]. Anion and cation states determine the hybridization matrix that is proportional to the wavefunction overlap of metal and Te holes, and Te p-like states penetrate Mn 4p orbitals. Indeed, in the vicinity of a Mn ion Te p-wave functions can be expanded in the basis of the Mn orbitals [32]. In the same vein, axial magnetic quadrupoles, formed by a correlation of anapole and orbital degrees of freedom, are expected to be different from zero. They are allowed in magnetic neutron scattering when the atomic wave function is drawn from two, or more, j-manifolds. Such is the case in the $j_{eff}$ = 1/2 state of Ir ions ($Ir^{4+}$, $5d^5$), often exploited in studies of $Sr_2IrO_4$, which is an admixture of manifolds with total angular momenta j = 3/2 and j = 5/2 [21].

## II. MAGNETIC PROPERTIES

The parent structure of $Mn_3Si_2Te_6$ is trigonal. It belongs to the space group $P\bar{3}1c$ (No. 163.79, BNS [33]) with Mn1 and Mn2 in Wyckoff sites 4f and 2c, respectively [2, 3]. Ferrimagnetic order depicted in Fig. 2 develops below ≈ 78 K without changes to the lattice structure [3]. Recently, the monoclinic magnetic space group $C2'/c'$ (No. 15.89, cell choice 1 [33]) with unique axis b was assigned to the magnetic order [5]. It belongs to the magnetic crystal class 2'/m' that is centrosymmetric, and permits ferromagnetism and a piezomagnetic effect, while a linear magnetoelectric effect is not allowed. The magnetic space group allows a ferromagnetic component within the (ac)-plane. This implies that application of a magnetic field along the c-axis, or any other direction within the (ac)-plane, does not change the C2′/c′ magnetic symmetry. A bulk magnetic moment is shown to be due to axial multipoles alone.

Mn1 and Mn2 use Wyckoff positions 8f (0, 1/3, ≈ 0) and 4e (0, 1/3, 1/4), respectively, in C2′/c′. Sites do not contain centres of inversion symmetry and both Mn ions are permitted axial (parity signature $\sigma_\pi = +1$) and polar (Dirac, $\sigma_\pi = -1$) magnetic multipoles. Monoclinic and hexagonal structures are related by $\mathbf{a} = -\mathbf{a}_h$, $\mathbf{b} = \mathbf{a}_h + 2\mathbf{b}_h = (0, a_o\sqrt{3}, 0)$ and $\mathbf{c} = -\mathbf{c}_h$, with hexagonal vectors $\mathbf{a}_h = (a_o, 0, 0)$, $\mathbf{b}_h = (1/2)(-a_o, a_o\sqrt{3}, 0)$ and $\mathbf{c}_h = (0, 0, c_o)$ [5]. Unit cell lengths $a \approx 7.029$ Å, $b \approx 12.175$ Å, $c \approx 14.255$ Å, and all cell angles = 90°. Miller indices for the monoclinic and hexagonal structures are denoted $(h, k, l)$ and $(H_o, K_o, L_o)$, respectively, with $h = -H_o$, $k = H_o + 2K_o$, $l = -L_o$. Atomic multipoles are calculated in orthogonal coordinates ($\xi$, $\eta$, $\zeta$) derived from $\mathbf{a}$, $\mathbf{b}$, $\mathbf{c}$, with unique axis $\mathbf{b}$ and $\eta$ parallel.

For our atomic description of electronic properties, Mn ions are assigned spherical multipoles $\langle O^K_Q \rangle$ with integer rank K and projections Q in the interval $-K \leq Q \leq K$. Cartesian and spherical components of a dipole $\mathbf{n} = (x, y, z)$, for example, are related by $x = (n_{-1} - n_{+1})/\sqrt{2}$, $y = i(n_{-1} + n_{+1})/\sqrt{2}$, $z = n_0$. A complex conjugate is defined as $\langle O^K_Q \rangle^* = (-1)^Q \langle O^K_{-Q} \rangle$, and a phase convention $\langle O^K_Q \rangle = [\langle O^K_Q \rangle' + i \langle O^K_Q \rangle'']$ for real and imaginary parts labelled by single and double primes, respectively. In which case, the diagonal multipole $\langle O^K_0 \rangle$ is purely real. Angular brackets $\langle ... \rangle$ denote the time-average, or expectation value, of the enclosed tensor operator, i.e., Mn multipoles feature in the electronic ground state of $Mn_3Si_2Te_6$.

Sites 8f occupied by Mn1 have no symmetry. They differ in this respect from sites 4e with symmetry $2_\eta'$, i.e., an anti-dyad parallel to the unique axis b of the monoclinic cell. In consequence, a multipole $\langle O^K_Q \rangle_2$ for Mn2 obeys the identity $\sigma_\theta (-1)^{K+Q} \langle O^K_{-Q} \rangle_2 = \langle O^K_Q \rangle_2$, where the time signature $\sigma_\theta = -1$ for a magnetic multipole. An alternative version of the identity for Mn2 multipoles is $\sigma_\theta (-1)^K \langle O^K_Q \rangle_2^* = \langle O^K_Q \rangle_2$. Thus, components of Mn2 magnetic dipoles parallel to the unique axis b are not permitted.

An electronic structure factor,

$$\Psi^K_Q = [\exp(i\boldsymbol{\kappa} \cdot \mathbf{d}) \langle O^K_Q \rangle_\mathbf{d}], \tag{1}$$

determines neutron and x-ray diffraction patterns for a particular space group [20]. In the present study, the implied sum in Eq. (1) is over positions $\mathbf{d}$ of Mn ions in a unit cell. The reflection vector $\boldsymbol{\kappa}$ with integer Miller indices referred to the monoclinic structure. There are three independent Te ions using Wyckoff positions 8f, and they possess general coordinates unlike Mn1. Tellurium nuclei contribute to neutron scattering for Miller indices that allow the scalar component of the electronic structure factor (K = 0, Q = 0) to be different from zero for $\sigma_\theta = +1$ (non-magnetic) and $\sigma_\pi = +1$ (axial). Examination of $\Psi^0_0(8f)$ with these auxiliary conditions reveals that $\Psi^0_0(8f) = 0$ for $(h, 0, l)$ with even $h$ and odd $l$, and $(0, k, l)$ with even $k$ and odd $l$; these are (basis-forbidden) reflections. Conventional neutron polarization analysis measures the ratio of magnetic and nuclear scattering amplitudes at basis-allowed, or core, Bragg reflections. In the case of x-rays, Thomson scattering from all ions is absent at forbidden reflections.

For Mn2 we find a structure factor,

$$\Psi^K_Q(2) = \langle O^K_Q \rangle_2 \exp(i\pi l/2)\,[1 + (-1)^{h+k}]\,[\exp(i\varphi) + \sigma_\pi (-1)^l \exp(-i\varphi)], \qquad (2)$$

with $\varphi = (2\pi k y)$ and $y = 1/3$ [5]. Bulk properties are found by setting $h = k = l = 0$. The corresponding structure factor can be different from zero for axial multipoles ($\sigma_\pi = +1$) and include dipoles ($K = 1$). C-centring is responsible for the reflection condition even ($h + k$). As we shall see, Mn1 multipoles share the latter condition and, also, those for a bulk magnetic moment. Multipoles with $\sigma_\theta (-1)^K = +1$ are purely real, i.e., magnetic multipoles with odd rank are purely real. In the present case of magnetic scattering, $\Psi^K_Q(2)$ for a reflection ($h, k, l$) is purely real for odd K, and even (odd) $l$ and $\sigma_\pi (-1)^l = +1\,(-1)$. Conventional neutron polarization analysis is available when magnetic and nuclear scattering have a like phase [35, 36, 37].

Specifically, reflections ($h, 0, l$) are absent for $\sigma_\pi (-1)^l = -1$. Bragg spots labelled by even $h$ and odd $l$ are caused by Dirac multipoles denoted $\langle G^K_Q \rangle$. Site symmetry allows anapoles $\langle G^1_\xi \rangle$ and $\langle G^1_\zeta \rangle$, cf. Fig. 1. Likewise, axial magnetic dipoles $\langle T^1_\xi \rangle$ and $\langle T^1_\zeta \rangle$ cause diffraction for ($h, 0, l$) with even $h, l$, since site symmetry for Mn2 does not include the parity signature.

Turning to Mn1,

$$\Psi^K_Q(1) = [1 + (-1)^{h+k}]\,[\exp(i\varphi) + \sigma_\pi \exp(-i\varphi)]$$
$$\times [\langle O^K_Q \rangle_1 + \sigma_\theta (-1)^{K+Q} (-1)^l \langle O^K_{-Q} \rangle_1], \qquad (3)$$

with $\varphi = (2\pi k/3)$. An alternative form of the the third factor $= [\langle O^K_Q \rangle_1 + \sigma_\theta (-1)^{K+l} \langle O^K_Q \rangle_1^*]$. A significant difference between Mn1 and Mn2 is that the former ion is permitted dipoles parallel to the unique axis b.

### III. NEUTRON DIFFRACTION

A magnetic scattering amplitude $\langle \mathbf{Q}_\perp \rangle$ generates an intensity $[|\langle \mathbf{Q}_\perp \rangle_1|^2 + |\langle \mathbf{Q}_\perp \rangle_2|^2]$ of unpolarized neutrons, where subscripts 1 and 2 denote independent ions Mn1 and Mn2. In more detail, $\langle \mathbf{Q}_\perp \rangle^{(\pm)} = [\mathbf{e} \times (\langle \mathbf{Q} \rangle^{(\pm)} \times \mathbf{e})]$ with a unit vector $\mathbf{e} = \boldsymbol{\kappa}/\kappa$, and superscripts refer to axial (+) and Dirac (−) multipoles. The intermediate amplitude $\langle \mathbf{Q} \rangle^{(+)}$ is proportional to the axial magnetic moment $\langle \boldsymbol{\mu} \rangle$ in the forward direction of scattering, with $\langle \mathbf{Q} \rangle^{(+)} = \langle \boldsymbol{\mu} \rangle/2$ for $\kappa = 0$, while the dipole in $\langle \mathbf{Q} \rangle^{(-)}$ can be related to anapoles [30, 38 - 42]. The magnetic content of a Bragg spot with overlapping nuclear and magnetic amplitudes can be detected by spin-flip scattering [22, 23, 43]. A fraction $\propto \{(1/2)\,(1 + P^2)\,|\langle \mathbf{Q}_\perp \rangle|^2 - |\mathbf{P} \cdot \langle \mathbf{Q}_\perp \rangle|^2\}$ of neutrons participate in events that change (flip) the neutron spin orientation, where $\mathbf{P}$ is the primary polarization. Taking $\mathbf{P} \cdot \mathbf{P} = 1$ yields a standard spin-flip signal [43],

$$(SF) = \{|\langle \mathbf{Q}_\perp \rangle|^2 - |\mathbf{P} \cdot \langle \mathbf{Q}_\perp \rangle|^2\}. \qquad (4)$$

In what follows, we show expressions for $\langle \mathbf{Q} \rangle$, or $\langle \mathbf{Q}_\perp \rangle$, appropriate to Bragg spots that expose intriguing facets of the magnetic structure assigned to $Mn_3Si_2Te_6$ by the monoclinic magnetic space group $C2'/c'$ [5].

Magnetic multipoles in neutron diffraction depend on the magnitude of the reflection vector, κ. An axial dipole $\langle \mathbf{T}^1 \rangle$ contains standard radial integrals $\langle j_0(\kappa) \rangle$ and $\langle j_2(\kappa) \rangle$ shown in Fig. 3, with $\langle j_0(0) \rangle = 1$ and $\langle j_2(0) \rangle = 0$. An approximation to the transition-metal dipole,

$$\langle \mathbf{T}^1 \rangle \approx (\langle \boldsymbol{\mu} \rangle/3) \, [\langle j_0(\kappa) \rangle + \langle j_2(\kappa) \rangle \, (g - 2)/g], \tag{5}$$

is often used [39, 40]. Here, the magnetic moment $\langle \boldsymbol{\mu} \rangle = g \langle \mathbf{S} \rangle$ and the orbital moment $\langle \mathbf{L} \rangle = (g - 2) \langle \mathbf{S} \rangle$. The coefficient of $\langle \mathbf{L} \rangle$ is approximate, while $\langle \mathbf{T}^1 \rangle = (1/3) \langle 2\mathbf{S} + \mathbf{L} \rangle$ for $\kappa \to 0$ is an exact result. This is the extent of the analysis used by Ye *et al.* to interpret their $Mn_3Si_2Te_6$ diffraction pattern [5]. Subtraction of patterns gathered above and below the magnetic ordering temperature (200 K and 5 K) provided an estimate of the magnetic content. Polarization analysis offers greater sensitivity to magnetic contributions. Higher order multipoles in the axial neutron scattering amplitude depend on the electronic position operator $\mathbf{n}$. The equivalent operator $[(\mathbf{S} \times \mathbf{n}) \, \mathbf{n}]$ for $\mathbf{T}^2$ shows that actually the quadrupole measures the correlation between the spin anapole $(\mathbf{S} \times \mathbf{n})$ and orbital degrees of freedom [21, 41]. The quadrupole $\langle \mathbf{T}^2 \rangle$ is proportional to $\langle j_2(\kappa) \rangle$ in Fig. 3.

The Dirac dipole $\langle \mathbf{d} \rangle$ depends on three radial integrals. We use [30, 41],

$$\langle \mathbf{d} \rangle = (1/2) \, [ \, i(g_1) \langle \mathbf{n} \rangle + 3 \, (h_1) \langle \mathbf{S} \times \mathbf{n} \rangle - (j_0) \langle \boldsymbol{\Omega} \rangle ]. \tag{6}$$

Radial integrals $(g_1)$ and $(j_0)$ shown in Fig. 3 diverge in the forward direction of scattering ($\kappa \to 0$). Not so for $(h_1)$ shown in Fig. 3, which also carries the κ-dependence of the Dirac quadrupole $\langle \mathbf{H}^2 \rangle \propto [(h_1) \langle \{\mathbf{S} \otimes \mathbf{n}\}^2 \rangle]$ observed in neutron diffraction from high-$T_c$ compounds Hg1201 and YBCO [22, 23]. Returning to Eq. (6), $\langle \boldsymbol{\Omega} \rangle = [\langle \mathbf{L} \times \mathbf{n} \rangle - \langle \mathbf{n} \times \mathbf{L} \rangle \,]$ is an orbital anapole (toroidal dipole) depicted in Fig. 1. We retain dipoles and quadrupoles in the following neutron diffraction amplitudes.

### A. Forbidden reflections

Ye *et al.* report the intensity of the Bragg spot $(0, 2, -1)$ for a sample temperature of 5 K as a function of a magnetic field applied along the crystal c-axis; Fig 3a in Ref. [5]. In the absence of multipoles other than dipoles $\langle \mathbf{Q} \rangle_1^{(+)} = (0, \langle Q_\eta \rangle_1^{(+)}, 0)$ with $\langle Q_\eta \rangle_1^{(+)} = [12 \cos(2\pi k/3) \langle T^1_\eta \rangle_1]$ for $(0, k, l)$ with even $k$ and odd $l$. Eq. (5) provides an estimate of the axial dipole. However, it is likely that quadrupoles $\langle \mathbf{T}^2 \rangle$ proportional to $\langle j_2(\kappa) \rangle$ contribute to the scattering amplitudes. In which case, $\langle Q_\xi \rangle_1^{(+)} \propto [e_\eta \, e_\zeta \, (\langle T^2_{+2} \rangle_1{}' + \sqrt{(3/2)} \langle T^2_0 \rangle_1)]$, $\langle Q_\zeta \rangle_1^{(+)} \propto [e_\eta \, e_\zeta \langle T^2_{+1} \rangle_1{}']$ and a contribution proportional to $[e_\zeta^2 \langle T^2_{+1} \rangle_1{}']$ is subtracted from $\langle Q_\eta \rangle_1^{(+)}$. Dirac multipoles are also allowed. Corresponding results for Mn2 ions are, $\langle \mathbf{Q} \rangle_2^{(+)} \propto [6 \sin(2\pi k/3) \, (\langle T^1_\xi \rangle_2, 0, \langle T^1_\zeta \rangle_2)]$ with quadrupoles $\langle T^2_{+1} \rangle_2{}''$ and $\langle T^2_{+2} \rangle_2{}''$ as corrections to dipoles in the (ac)-plane and $\langle Q_\eta \rangle_2^{(+)} \propto [e_\eta \, e_\zeta \langle T^2_{+2} \rangle_2{}'']$. Reflections of the type $(h, 0, l)$ with even $h$ and odd $l$ are more selective than $(0, k, l)$ with even $k$ and odd $l$, since Mn1 Dirac multipoles and Mn2 axial multipoles are not permitted for $k = 0$.

For Mn1 and $(h, 0, l)$ with even $h$ and odd $l$ we find $\langle Q_{\perp\xi} \rangle_1^{(+)} = \langle Q_{\perp\zeta} \rangle_1^{(+)} = 0$ and,

$$\langle Q_{\perp\eta}\rangle_1^{(+)} \approx 8\,[(3/2)\,\langle T^1_\eta\rangle_1 + \sqrt{3}\{e_\xi e_\zeta(\langle T^2_{+2}\rangle_1' - \sqrt{(3/2)}\,\langle T^2_0\rangle_1) + (e_\xi^2 - e_\zeta^2)\,\langle T^2_{+1}\rangle_1'\}]. \quad (7)$$

Unit vectors are,

$$e_\xi = h/[h^2 + (rl)^2]^{1/2}, \quad e_\zeta = rl/[h^2 + (rl)^2]^{1/2} \text{ with } r = (a/c) \approx 0.493. \quad (8)$$

We find $\langle Q_{\perp\xi}\rangle_2^{(-)} = \langle Q_{\perp\zeta}\rangle_2^{(-)} = 0$ and,

$$\langle Q_{\perp\eta}\rangle_2^{(-)} \approx 4\,(-1)^m\,[e_\xi\,\langle d_\zeta\rangle_2 - e_\zeta\,\langle d_\xi\rangle_2 + (3/\sqrt{5})\,(e_\zeta\,\langle H^2_{+1}\rangle_2'' - e_\xi\,\langle H^2_{+2}\rangle_2'')]. \quad (9)$$

with Miller index $l = (2m + 1)$. Axial and Dirac contributions in Eqs. (7) and (9) are distinguished by different dependences on the reflection vector. Both amplitudes are purely real. However, conventional neutron polarization analysis is not available in the absence of nuclear scattering [35, 36, 37].

### B. Allowed reflections

Axial amplitudes for Mn1 and reflections $(h, k, l)$ with even $h + k$, $l$ are,

$$\langle Q_\xi\rangle_1^{(+)} \approx (3/2)\,\langle T^1_\xi\rangle_1 - \sqrt{3}[e_\xi e_\zeta\,\langle T^2_{+2}\rangle_1'' + (e_\eta^2 - e_\zeta^2)\,\langle T^2_{+1}\rangle_1''],$$

$$\langle Q_\eta\rangle_1^{(+)} \approx \sqrt{3}\,e_\eta\,[e_\xi\,\langle T^2_{+1}\rangle_1'' + e_\zeta\,\langle T^2_{+2}\rangle_1''], \quad (10)$$

$$\langle Q_\zeta\rangle_1^{(+)} \approx (3/2)\,\langle T^1_\zeta\rangle_1 - \sqrt{3}[e_\xi e_\zeta\,\langle T^2_{+1}\rangle_1'' + (e_\eta^2 - e_\xi^2)\,\langle T^2_{+2}\rangle_1''].$$

Dipole components in the $\xi$-$\zeta$ plane contribute. A common factor $8\cos(2\pi k/3)$ is omitted in Eq. (10). Recall that $e_\xi \propto h$, $e_\eta \propto k$ and $e_\zeta \propto l$. Dirac Mn1 amplitudes have a common factor $[8\sin(2\pi k/3)]$, and they are zero for the special case $(h, 0, l)$. We find,

$$\langle Q_{\perp\xi}\rangle_1^{(-)} \approx -e_\eta\,[\langle d_\zeta\rangle_1 + (3/\sqrt{5})\,\{(1 - 2e_\xi^2)\,\langle H^2_{+2}\rangle_1'' + 2e_\xi e_\zeta\,\langle H^2_{+1}\rangle_1''\}],$$

$$\langle Q_{\perp\eta}\rangle_1^{(-)} \approx e_\xi\,\langle d_\zeta\rangle_1 - e_\zeta\,\langle d_\xi\rangle_1 + (3/\sqrt{5})\,(1 - 2e_\eta^2)\,[e_\zeta\,\langle H^2_{+1}\rangle_1'' - e_\xi\,\langle H^2_{+2}\rangle_1''],$$

$$\langle Q_{\perp\zeta}\rangle_1^{(-)} \approx e_\eta\,[\langle d_\xi\rangle_1 + (3/\sqrt{5})\,\{(1 - 2e_\zeta^2)\,\langle H^2_{+1}\rangle_1'' + 2e_\xi e_\zeta\,\langle H^2_{+2}\rangle_1''\}]. \quad (11)$$

Amplitudes for Mn1 and Mn2 are similar, because Miller index $l$ is even.

### IV. RESONANT X-RAY DIFFRACTION

The Bragg angle $\theta$ in Fig. 4 for the reflection $(h, 0, l)$ is determined by,

$$\sin(\theta) = (\lambda/2a)\,[h^2 + (rl)^2]^{1/2}, \quad (12)$$

with a wavelength $\lambda \approx (12.4/E)$ Å, and the resonance energy E in keV. Manganese absorption edges used in an electric dipole- electric dipole (E1-E1) scattering event have energies $E \approx 6.537$ keV for the K edge (1s → 4p), and $L_2 \approx 0.649$ keV and $L_3 \approx 0.638$ keV (2p → 3d). Enhancement at the K edge reveals Mn orbital states alone [19]. Previous studies using resonant x-ray diffraction at the K-edge of 3d-transition metal compounds include $V_2O_3$, $\alpha$-$Fe_2O_3$ and NiO [11, 15, 45, 46]. Spin and orbital states are revealed at L edges and their contributions to absorption satisfy useful sum-rules [47]. The space group forbidden Bragg spot (0, 0, 1) can be accessed at L edges, while many Bragg spots indexed by $(h, 0, l)$ with even $h$ are available using the K edge. Radial matrix elements for Mn ($3d^5$) calculated from Cowan's program [44] are (1s|R|4p)/$a_o$ = − 0.00354 and (2p|R|3d)/(1s|R|4p) = 58.25, where $a_o$ is the Bohr radius. Dirac multipoles contribute to an E1-E2 event allowed by 3d-4p mixing (1s → 3d, 1s → 4p), and the Mn E2 radial integral (1s|$R^2$|3d)/ $a_o^2$ = 0.00095. An E1-E2 event manifests itself as a pre-edge feature to K edge absorption spectra [48, 49].

Four states of polarization in the primary (σ, π) and secondary (σ′, π′) x-ray beams are depicted in Fig. 4. The scattering amplitude for primary σ and secondary π′ is denoted (π′σ), for example. Intensity of a Bragg spot in the rotated channel of polarization is proportional to |(π′σ)|$^2$, and likewise for unrotated channels. Proportionality factors include radial integrals, and E1-E1 and E2-E2 amplitudes for the K edge are proportional to (1s|R|4p)$^2$ and [(1s|$R^2$|3d)/λ]$^2$, respectively. Reported amplitudes are functions of the angle ψ that measures rotation of the illuminated crystal about the reflection vector and they are derived from universal expressions [20, 50].

It follows from the structure factor Eq. (3) for Mn1 that diffraction indexed (*h*, 0, *l*) proceeds by a parity-even absorption event E1-E1, or E2-E2, with σ$_θ$ (−1)$^K$ = +1 [20, 50]. On the other hand, selecting forbidden reflections with even *h* and odd *l* means diffraction by Mn2 proceeds by a parity-odd event E1-E2. Consider first Mn1, an E1-E1 event and odd *l*. The corresponding electronic structure factor Ψ$^K_Q$(1) is purely imaginary. An axial dipole ⟨$t^1_η$⟩$_1$ contributes to both (π′π) and (π′σ), and not (σ′σ) [20, 50]. For the E1-E1 rotated channel,

$$(\pi'\sigma) = \cos(\theta) \cos(\psi) (i/\sqrt{2}) \langle t^1_\eta \rangle_1 - \sin(\theta) \cos(2\psi) [\alpha \langle t^2_{+1} \rangle_1'' + \beta \langle t^2_{+2} \rangle_1'' ]$$

$$+ \cos(\theta) \sin(\psi) [\beta \langle t^2_{+1} \rangle_1'' - \alpha \langle t^2_{+2} \rangle_1'' ]. \quad (13)$$

The reciprocal lattice vector **b*** and the η-axis are antiparallel at the start of an azimuthal angle scan (ψ = 0). In Eq. (13), α = cos(χ), β = sin(χ) with,

$$\cos(\chi) = h/[h^2 + (rl)^2]^{1/2}, \quad (14)$$

and we omit a common factor = 8. Schmitt *et al*. [51] set out attributes of (π′σ) that make it attractive to measurements. Quadrupoles ⟨$t^2_Q$⟩$_1$″ in Eq. (13) are time-even (non-magnetic) and account for T & T scattering, i.e., nominally weak scattering created by angular anisotropy in the electronic charge distribution of the resonant ion [24, 25].

Intensity picked out by circular polarization in the primary photon beam = $P_2$Υ where [16, 17, 18, 52],

$$\Upsilon = \{(\sigma'\pi)^*(\sigma'\sigma) + (\pi'\pi)^*(\pi'\sigma)\}''. \quad (15)$$

The Stokes parameter $P_2$ (a purely real pseudoscalar) measures helicity in the primary x-ray beam. Since intensity is a true scalar, Υ and $P_2$ must possess identical discrete symmetries, specifically, both scalars are time-even and parity-odd. The signature is extracted from observed Bragg spot intensities by subtraction of intensities measured with left- and right-handed primary x-rays, or x-rays with opposite helicities, namely, ± $P_2$. From Eq. (15), ions

Mn1 observed at ($h$, 0, $l$) with even $h$ and odd $l$ using an E1-E1 absorption event possess a chiral signature,

$$\Upsilon = - (1/\sqrt{2}) \langle t^1_\eta \rangle_1 \cos(\theta) \sin(\psi) \{\sin(2\theta) \sin(\psi) [\beta \langle t^2_{+1} \rangle_1'' - \alpha \langle t^2_{+2} \rangle_1'' ]$$
$$+ 2 [1 - (\cos(\theta) \sin(\psi))^2] [\alpha \langle t^2_{+1} \rangle_1'' + \beta \langle t^2_{+2} \rangle_1'' ]\}, \qquad (16)$$

apart from a factor = 64. A non-zero $\Upsilon$ relies on interference between a magnetic dipole $\langle t^1_\eta \rangle_1$ parallel to the unique axis b and charge-like quadrupoles $\langle t^2_Q \rangle_1''$. Estimates of the two quadrupoles can be inferred from contributions to $\Upsilon$ that are even and odd functions of $\psi$.

Continuing with ($h$, 0, $l$) with even $h$ and odd $l$, Mn2 ions diffract Dirac multipoles including anapoles $\langle G^1_\xi \rangle_2$ and $\langle G^1_\zeta \rangle_2$. For an E1-E2 event,

$$(\sigma'\sigma) \approx (4i/5) \sqrt{3} (-1)^m \cos(\theta) \sin(\psi) [\beta \langle G^1_\xi \rangle_2 - \alpha \langle G^1_\zeta \rangle_2]. \qquad (17)$$

at the level of anapoles. A complete result for $(\sigma'\sigma)$ appears in an Appendix. Notably, a simple dependence on the azimuthal angle $\sin(\psi)$ extends to quadrupoles in the unrotated amplitude. The corresponding chiral signature is zero, because all four diffraction amplitudes in Eq. (15) are purely imaginary. Indeed, $(\sigma'\sigma)$ and $(\pi'\pi)$ have identical structures for an E1-E2 event whereas they differ by a dipole contribution for an E1-E1 event [50].

### V. CONCLUSIONS

In summary, we have shown through calculations informed by magnetic symmetry that ferrimagnetic $Mn_3Si_2Te_6$ likely possesses many unusual properties, in addition to a recently discovered unique CMR [6, 7]. Our revelations are derived from a monoclinic space group previously proposed on the basis of an investigation using magnetic neutron diffraction [5]. Manganese ions use two symmetry independent sites, as in the trigonal paramagnetic structure [1]. Beyond axial magnetic dipoles, depicted in Fig. 2, are magnetic quadrupoles and anapoles that are possibly visible in an analysis of high-resolution diffraction patterns. By way of an example, the intensity of a Bragg spot with a reflection vector $\kappa \approx 1.122$ Å$^{-1}$ is reported in Ref. [5]. Contributions to the neutron scattering amplitude from spin and orbital anapoles are accompanied by atomic form factors $(h_1)$ and $(j_0)$ with values $\approx 10\%$ of the axial form factor $\langle j_0 \rangle$ for this reflection vector, as can be seen by inspection of Fig. 3. In consequence, the spin anapole might account for $\approx 30\%$ of the neutron scattering amplitude Eq. (6). Magnetic quadrupoles have form factors $\langle j_2 \rangle$ that peak above $\approx 7$ Å$^{-1}$.

A nominally weak Bragg spot, forbidden by reflection conditions for the parent structure of $Mn_3Si_2Te_6$, has been measured as a function of a magnetic field [5]. The standard method to estimate the magnetic content of diffraction patterns is to analyse the difference between two measured above and below the magnetic ordering temperature. Polarization analysis in neutron diffraction provides greater sensitivity to the magnetic content of Bragg spots with overlapping magnetic and nuclear contents permitted by reflection conditions of the parent structure [21, 35-37, 43]. In presenting our results for future neutron and resonant x-ray diffraction experiments, we pay particular attention to weak magnetic Bragg spots. In x-ray diffraction by non-magnetic materials they arise from Templeton-Templeton scattering created from angular anisotropy in electronic charge distributions [24, 25].

Magnetic multipoles revealed by resonant x-ray diffraction do not contain atomic form factors [14, 15, 20]. Axial and Dirac multipoles contribute to different diffraction amplitudes at different resonance energies. Orbital degrees of freedom in the Mn p-type valence state are exposed through signal enhancement by an (parity-even) electric dipole - electric dipole (E1-E1) event at the Mn K edge [19, 20]. Absorption at an L edge samples the Mn d-type valence state. Whereas, Dirac multipoles are exposed in an (parity-odd) electric dipole - electric quadrupole (E1-E2) event using a pre-edge feature to the Mn K edge [48, 49]. Amplitudes in Section IV are functions of rotation of the $Mn_3Si_2Te_6$ sample about the reflection vector, and azimuthal-angle displays are different for different absorption events. Notably, the more abundant Mn ions possess an E1-E1 chiral signature Eq. (16) not found for the second type of Mn ions for the same diffraction conditions.

**ACKNOWLEDGEMENT** 

## APPENDIX

Mn2, reflection ($h$, 0, $l$) even $h$ and $l = (2m + 1)$, a common factor 4, and an E1-E2 absorption event. The abbreviations $\alpha = \cos(\chi)$, $\beta = \sin(\chi)$, with $\cos(\chi) = h/[h^2 + (rl)^2]^{1/2}$ with $r = (a/c) \approx 0.493$.

$$(\sigma'\sigma) = (4i/5)\sqrt{3}\,(-1)^m\,\cos(\theta)\sin(\psi)\,\{\alpha\,[-\langle G^1_\zeta\rangle_2 + (1/3)\sqrt{10}\,\langle G^2_{+2}\rangle_2'']$$
$$+ \beta\,[\langle G^1_\xi\rangle_2 - (1/3)\sqrt{10}\,\langle G^2_{+1}\rangle_2''] + \sqrt{(2/3)}\,\alpha\,[1 - 5\alpha^2\cos^2(\psi)]\,\langle G^3_0\rangle_2$$
$$+ (1/3)\sqrt{2}\,\beta\,[1 - 15\alpha^2\cos^2(\psi)]\,\langle G^3_{+1}\rangle_2' + (2/3)\sqrt{5}\,\alpha\,[1 - 3(1 + \beta^2)\cos^2(\psi)]\,\langle G^3_{+2}\rangle_2'$$
$$- \sqrt{(10/3)}\,\beta\,[\alpha^2 + (1 - 4\alpha^2)\cos^2(\psi)]\,\langle G^3_{+3}\rangle_2'\}. \quad\quad\quad (A1)$$

Evidently, octupole contributions to $(\sigma'\sigma)$ modify the simple $\sin(\psi)$ azimuthal-angle dependence that hallmarks anapoles and quadrupoles.

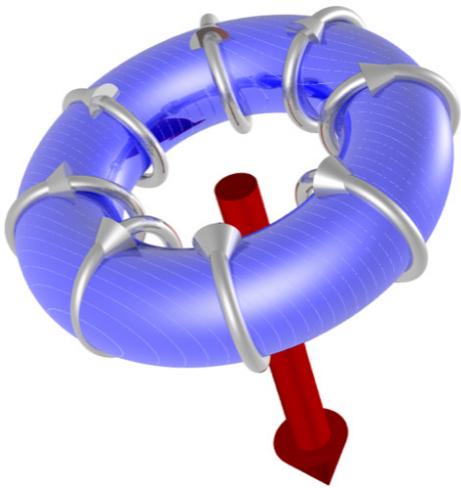

FIG. 1. Depiction of a toroidal dipole, also known as an anapole. Figure prepared by V. Scagnoli.

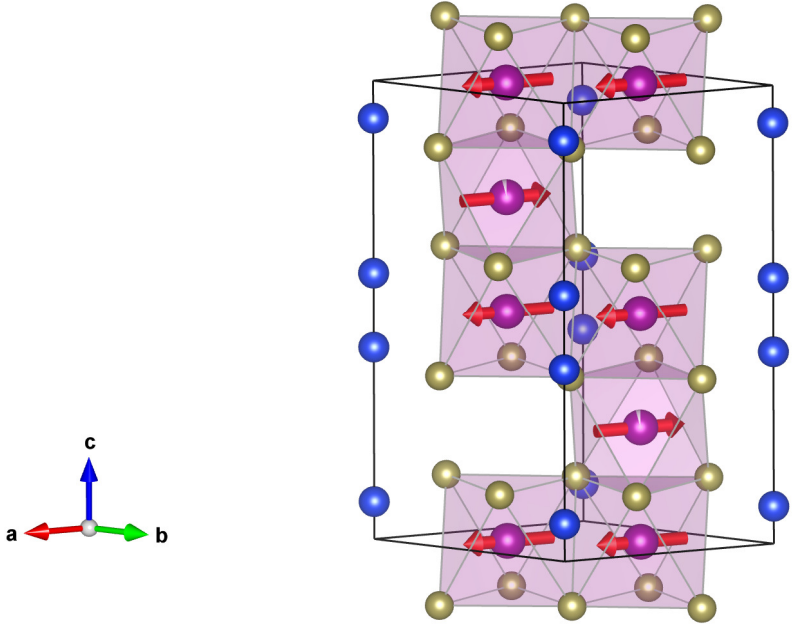

FIG. 2. Mn axial dipoles in $Mn_3Si_2Te_6$ depicted in an hexagonal setting. Si ions in blue and $MnTe_6$ units. After A. F. May *et al.* [3]. Reproduced from MAGNDATA [34].

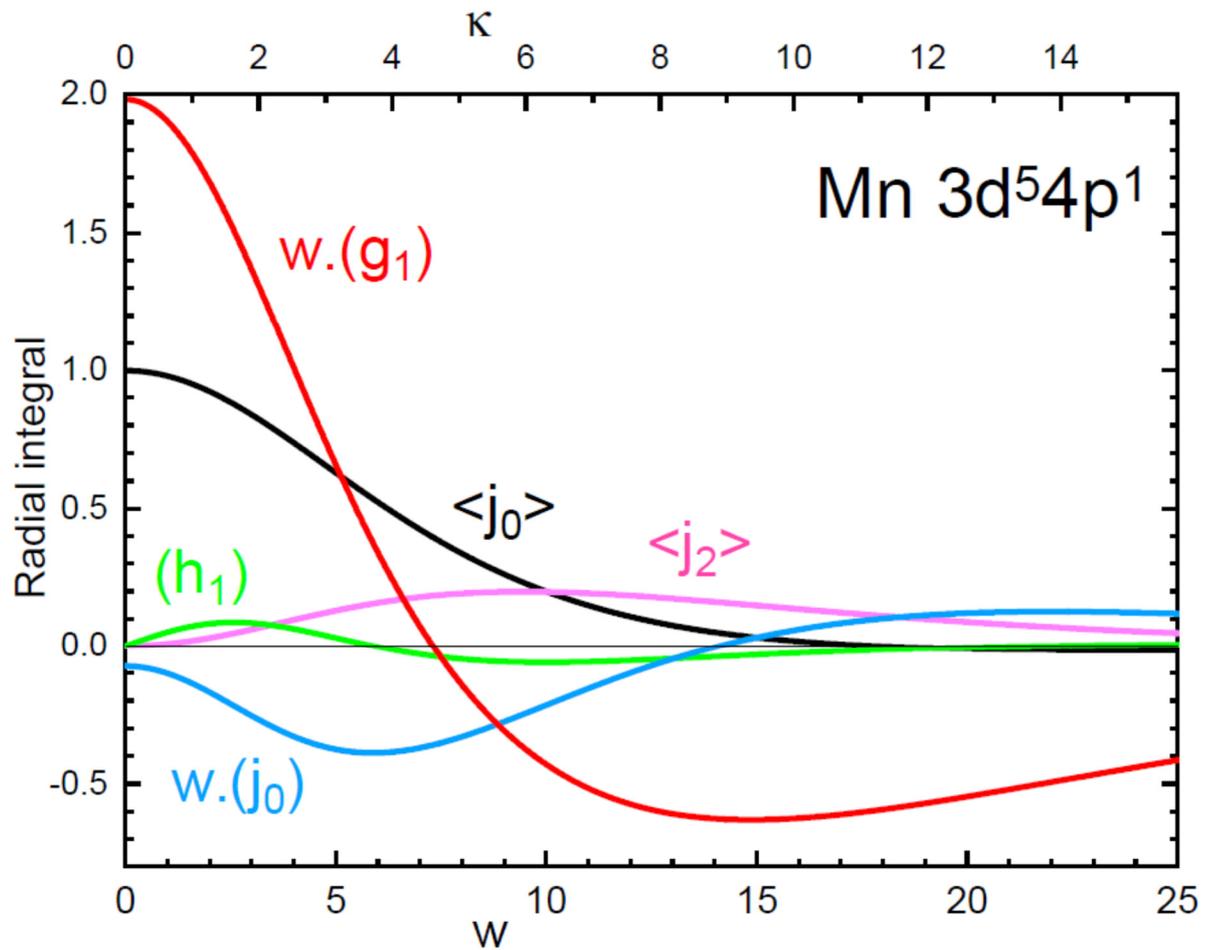

FIG. 3. Radial integrals for $Mn^{2+}$ ($3d^5$) displayed as a function of the magnitude of the reflection vector $\kappa = 4\pi s$ with $s = \sin(\theta)/\lambda$ (Å$^{-1}$), Bragg angle $\theta$ and neutron wavelength $\lambda$.

Also, a dimensionless variable w = $3a_o\kappa$ where $a_o$ is the Bohr radius. Blue and purple lines are standard radial integrals $\langle j_0(\kappa)\rangle$ and $\langle j_2(\kappa)\rangle$ that occur in the axial dipole Eq. (5). Red, green and blue curves are radial integrals in the polar dipole Eq. (6). Two integrals ($g_1$) and ($j_0$) diverge in the forward direction of scattering, and quantities w($g_1$) and w($j_0$) are displayed for this reason . Calculations, performed with Cowan's atomic code [44], and figure made by G. van der Laan.

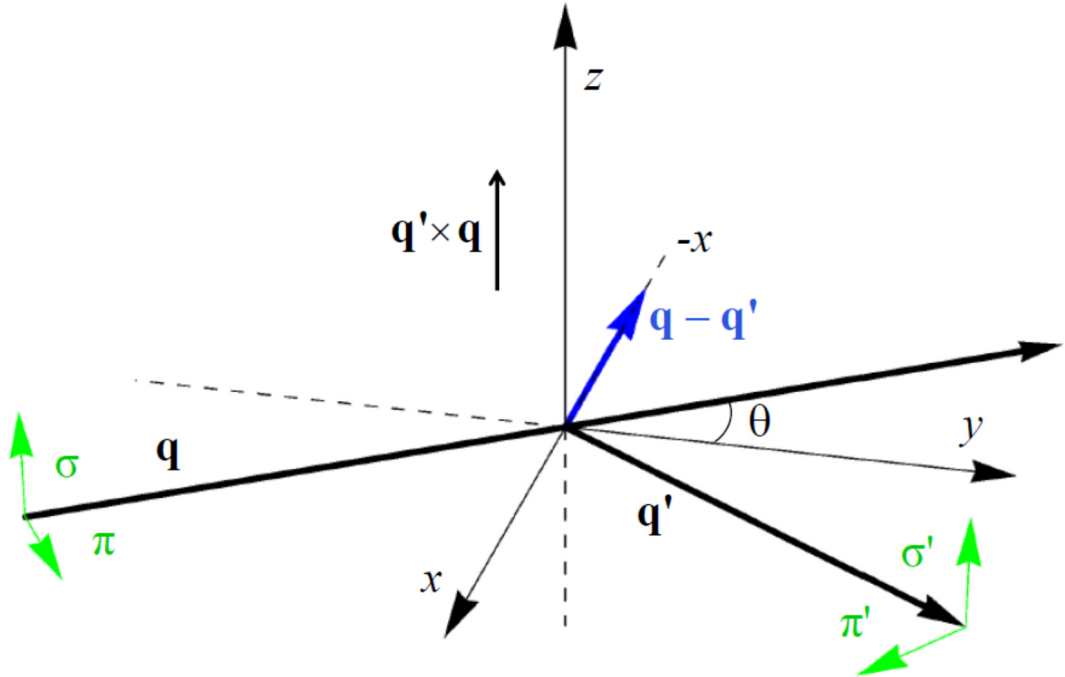

FIG. 4. Primary (σ, π) and secondary (σ', π') states of polarization. Corresponding wavevectors **q** and **q'** subtend an angle 2θ specified in Eq. (12). The Bragg condition for diffraction is met when **q** − **q'** coincides with the reflection vector indexed (h, k, l). Monoclinic crystal vectors **a**, **b**, and **c** that define local axes (ξ, η, ζ) and the depicted Cartesian (x, y, z) coincide in the nominal setting of the crystal where the azimuthal angle ψ = 0.